\documentclass[reprint]{revtex4-1}

\usepackage{txfonts}
\usepackage{graphicx}
\begin{document}
\title{The effect of laziness in chasers in group chase and escape model}
\author{Makoto Masuko}
\email{masuko@cmr.t.u-tokyo.ac.jp}
\author{Takayuki Hiraoka}
\author{Nobuyasu Ito}
\author{Takashi Shimada}
\email{shimada@ap.t.u-tokyo.ac.jp}
\affiliation{Department of Applied Physics, Graduate School of Engineering, University of Tokyo, Japan}

\begin{abstract}
The effect of laziness in the group chase and escape problem is studied using a simple model.
The laziness is introduced as random walks in two ways: uniformly and in a ``division of labor” way.
It is shown that, while the former is always ineffective, the latter can improve the efficiency of catching,
through the formation of pincer attack configuration by diligent and lazy chasers.
\end{abstract}

\maketitle
A recently proposed simple model of multiple chasers and multiple escaping targets\cite{Kamimura} shed a new light on a classical chase and escape problem.
In their group chase and escape model, each chaser approaches its nearest target and each target leaves from the nearest chaser.
In spite of its simplicity, non-trivial cooperative chasing behavior emerges.
While such cooperative dynamics allow chasers to catch targets escaping with the same velocity,
even a group of chasers sometimes need a very long duration until catching the entire targets. 
Their results imply that such cases stem from the situation that all the chasers follow a same target as a single cluster, because of thir diligent nature of chasing the nearest target.

On the other hand, the effect of laziness in animal groups has been investigated.
Previous studies revealed that appropriate laziness may improve the efficiency or stability of a whole group \cite{Hasegawa, Nishimori}.
Although group chase and escape model is one of the models of group strategies
and various extensions of the original model are suggested and studied, the effect of laziness on group chase and escape is not known.
We here extend the model by introducing lazy chasers and study its effect on the catching efficiency.

\begin{figure}[t]
\centering
\includegraphics[width = 7cm, bb= 0 0 1331 1041]{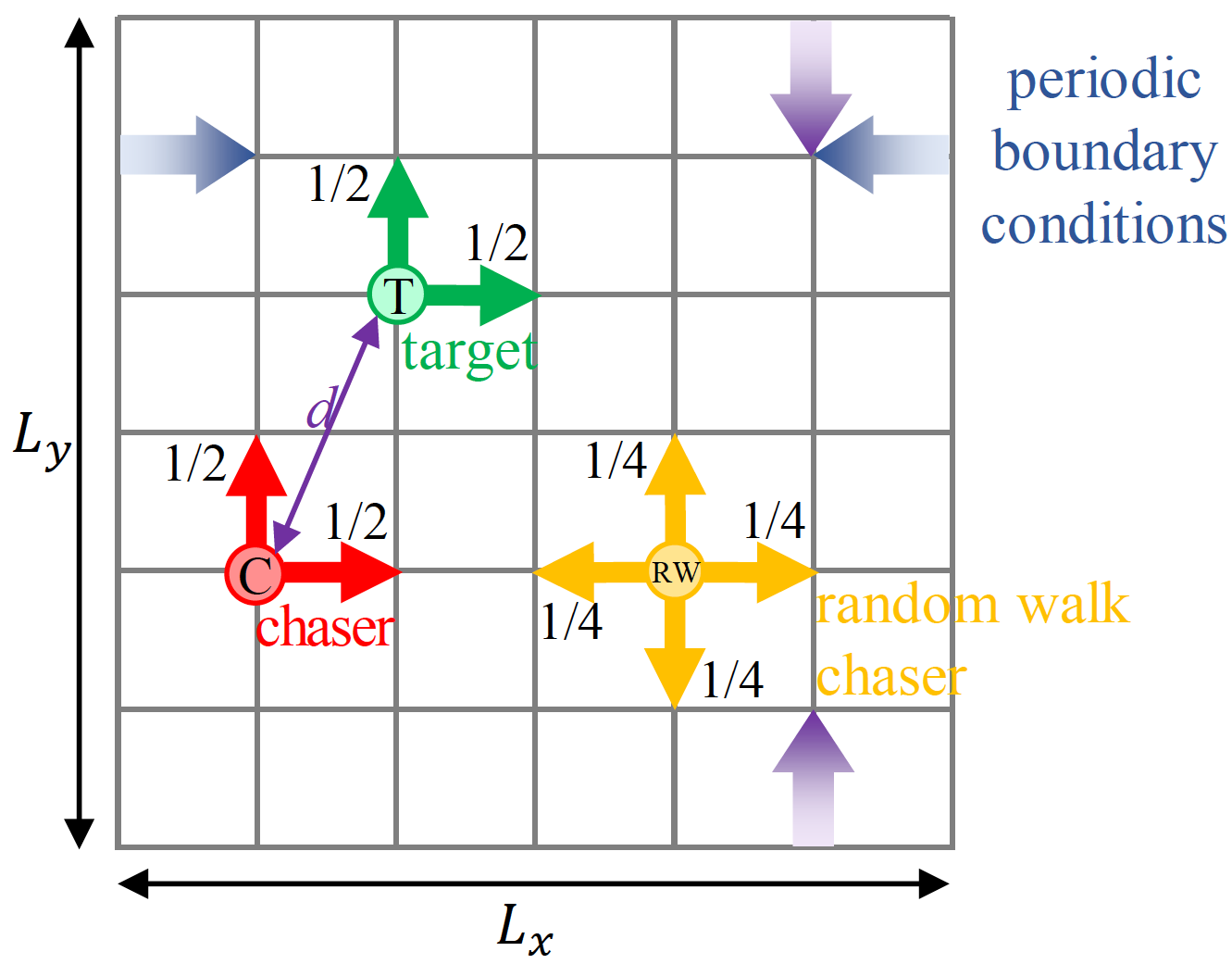}
\caption{
The schematic diagram of the group chase and escape model with lazy (random walk) chaser.
}
\label{fig:group_cae_model}
\end{figure}

As shown in Fig. \ref{fig:group_cae_model}, we consider a square lattice with periodic boundaries.
Each vertex can be occupied by only one animal, a chaser or a target.
Initially, $N_C$ chasers and $N_T$ targets are randomly distributed on the lattice.
At each time-step, chasers and targets determine the directions to move, according to the following rule.
Each chaser first searches for the nearest target, according to the geometric distance
$d=\sqrt{(x_C+x_T)^2+(y_C+y_T)^2}$, where $(x_C, y_C)$ and $(x_T, y_T)$
denote the positions of the chaser and the target, respectively.
Then the chaser selects one direction so that the move toward it decreases $d$.
In general, there can be two possible directions.
In that case, one is chosen with equal probability $1/2$.
Targets also decide the directions to move in the same way, to increase the distance from the nearest chaser.
After determining the directions to move for all the animals,
we first move chasers by one step with randomly determined order at each time-step.
Targets are next moved by one step in random order.
If the vertex to which a chaser intends to move is occupied by another chaser, the chaser does not move at that time-step.
Targets cannot move to occupied vertices, similarly.
When a chaser moves to the vertex on which a target exists, the target is caught by the chaser and eliminated from the system. 
The chasing and escaping process as above is repeated until either all targets are caught or time reaches a certain value to abort calculation.
$T$ is defined as the time needed for catching all targets, and the efficiency of catching is evaluated by the shortness of its ensemble average $\langle T \rangle$.

To model a lazy chaser which does not chase the targets diligently,
we substitute a random walk with equal probability $1/4$ for each direction for the diligent chasing.
We consider two ways of bringing laziness to a group of chasers: either uniformly or in a “division of labor” way (non-uniformly).
In uniform laziness case, each chaser may perform a random walk with certain probability $P\: (0\le P\le 1)$ at each time-step.
The probability of random walk $P$ is common to all chasers.
In division of labor case, in contrast, $N_{lazy} = RN_C$ chasers always do a random walk and the other $(1-R)N_C$ chasers always chase the nearest target.
Here, $R\: (0\le R\le 1)$ denotes the ratio of random walk chasers.
Note that $P$ and $R$ are the parameter of average laziness, because those represent the ratio between diligent and lazy (random) actions. 

We first simulate the model with uniform laziness.
In this case, the average of the duration until total catch $\langle T \rangle$ for $P>0$ is found to be always longer than that for the case of $P=0$ (not shown).
Therefore, it is concluded that uniform laziness is less effective than original diligent group chasing in this model.

\begin{figure}[t]
\centering
\includegraphics[width = 0.9\hsize]{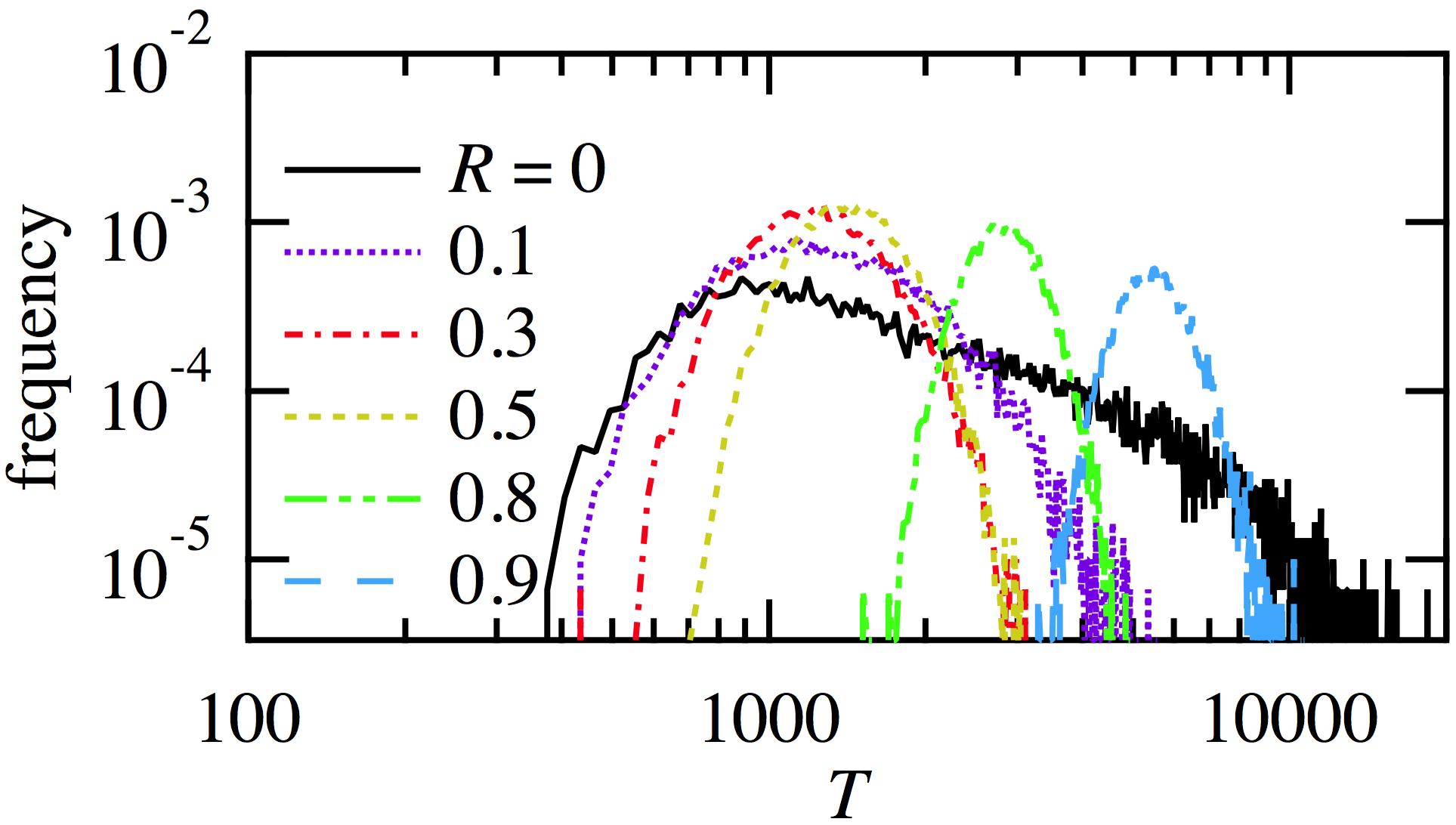}

\caption{
The normalized frequency distributions of $T$ for different $R$. The each distribution is obtained from $10,000$ initial configurations.
}
\label{fig:dist_T}
\end{figure}

\begin{figure}[t!]
\centering	
\includegraphics[width = 0.9\hsize]{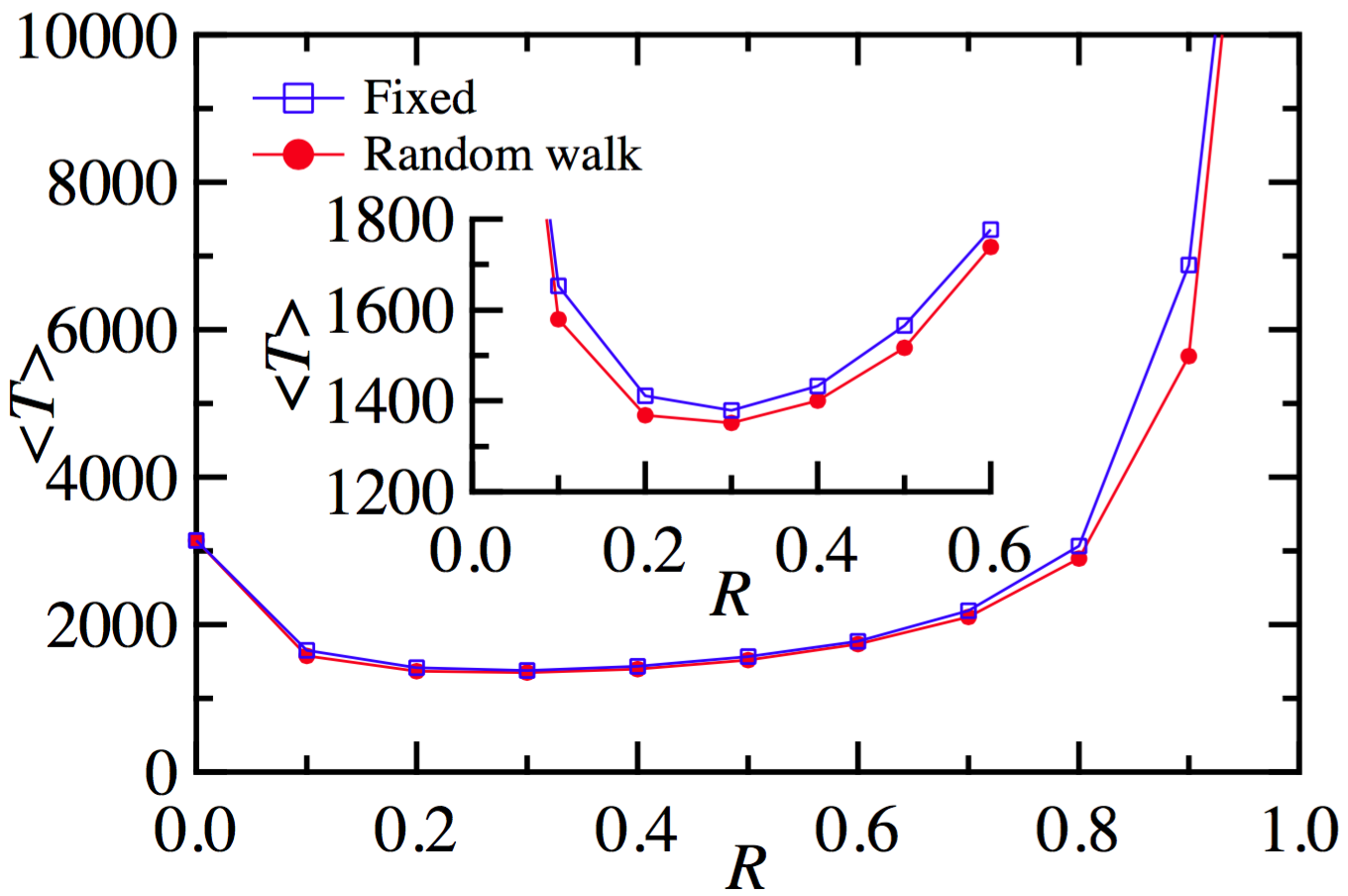}
\caption{
$\langle T \rangle$ as a function of the ratio of lazy chasers $R$, introduced as a random walkers (red) and fixed chasers (blue).
The inset is the enlargement in the range of $0\le R \le 0.6$.
The parameters are the same as those in Fig. \ref{fig:dist_T}.
The average for each $R$ is taken from $10,000$ samples and the error bars are smaller than the symbol sizes.
}
\label{fig:RWandFixed}
\end{figure}

In the ``division of labor'' case, however, we find a different behavior.
Simulation is carried out for $R=0,0.1,0.2,\ldots ,1$ with keeping other system parameters at
$L_x=L_y=200$ and $N_C=N_T=40$.
For each $R$, we perform simulations from $10,000$ independent initial conditions.
The frequency distributions of the time until total catch ($T$) for each $R$ is shown in Fig. \ref{fig:dist_T}.
As shown in the figure, the frequency distribution for $R = 0$ (black line) is asymmetric and has a broad tail
which reaches over $T=10,000$, while its peak is at around $1,000$.
It means that, although catching all targets needs only approximately $1,000$ time-steps in most cases,
it may take several to tens times more of time-steps in some cases.
By comparing the distribution for $R > 0$ with that for $R = 0$,
one can see that replacing some chasers with random walkers suppresses the tail of distribution drastically and makes unlucky cases rarer, with slight increase in the peak position.
And this change in $T$ distribution results in the reduction in its average, $\langle T \rangle$ (Fig. \ref{fig:RWandFixed}).
Additional simulation for some different system sizes with fixing the density of both chasers and targets at the same value, $0.001$
confirms that the positive effect of $R$ on $\langle T \rangle$ is robust among the different system size (not shown).
It indicates that the effect of random walk chasers depends on not the number of them but the spacial density of them.

To understand the drastic effect of introducing the lazy chasers,
let us first review the cause of the broad tail of $T$ distribution from the characteristics of group chasing.
In the present group chase and escape model, single chaser can never catch up with target, because the speed of the chasers is equal to that of the targets.
However, when some chasers follow the same target from different directions, they may catch the target by surrounding it and intercepting its escape route.
In such a ``pincer attack'' configuration, chasers behave as if they cooperate with each other to catch the target.
In contrast to such cooperative catching, the clustering of chasers makes $T$ longer.
When some chasers recognize the same target as the nearest one and follow it from the same direction, they may form a cluster.
Once a cluster of chasers is formed, clustering chasers rarely scatter because each chaser in the same cluster almost always intends to follow the same target and thus move in the same direction.
Moreover, clusters may fuse in time and form a larger cluster.
When almost all chasers form single cluster, chasing and escaping is like ``cat-and-mouse game''  due to periodic boundary conditions.
In such a case, catching all targets takes very long time.
These two effects can explain the peak around $T \sim 1,000$ (cooperative chase) and the asymmetry brought by the broad tail (``cat-and-mouse'' game)
of the frequency distribution of $T$ of the original model ($R = 0$ in Fig. \ref{fig:dist_T}).

One can see in Fig. \ref{fig:dist_T} that replacing some chasers with random walkers decreases the probability of unlucky cases,
although random walk chasers do not directly prevent the clustering.
The contribution of lazy chasers is from the fact that targets cannot distinguish a lazy chaser from diligent chasers.
In division of labor case, even if almost all diligent chasers form a cluster and ``cat-and-mouse game" situation emerges,
accidental chasing in a direction of random walk chaser enables chasers’ cluster to reduce the distance to the target
when the target recognizes random walk chaser as the nearest one and try to escape from that, not from the cluster.
Therefore, doping some lazy chasers is expected to destroy unlucky chasing configurations and suppresses the tail of frequency distribution of $T$.
The fact that uniform probabilistic laziness is always ineffective means that the negative effect of the decrease in the average chasing speed is greater than the positive effect of preventing chasers from forming a compact cluster.

To investigate the mechanism of the positive contribution of random walkers in more detail,
we next replace the $RN_C$ lazy chasers are by fixed chasers, which do not move at all.
As shown in Fig. \ref{fig:RWandFixed}, the behaviors of $\langle T \rangle$ of this fixed chaser model and the random walker model are almost the same,
although $\langle T \rangle$ of the fixed lazy chaser model is always larger than that of random walker model.
It implies that the contribution of random walk chasers to group chase is mainly from its ability to form a ``pincer attack'' configuration.
The improvement in the efficiency by random walk chasers is considered to be from the two effects.
Random walk chasers effectively occupy larger area and limit the area for targets to escape better than fixed chasers do.
Another reason is that, when the lazy chasers are distributed in a special initial configuration,
such as gathering in a certain region or giving a path for targets,
fixed chasers cannot contribute to the formation of ``pincer attack'' configuration while the random walk chasers will dissolve their bad initial configuration.

We have shown that replacing a portion of diligent chasers by lazy chasers generally improves the efficiency of catching,
while the introduction of laziness uniformly to the all chasers is always ineffective.
Our findings of the positive effect of introducing ``lazy'' members, especially in a ``division of labor'' way,
is similar to the previously other animal behaviors\cite{Hasegawa, Nishimori}.

\vspace{3mm}
We thank Toru Ohira and Hiraku Nishimori for discussion. This work was supported by CREST,
JST and JPS-NRF bilateral project.

\end{document}